# The relationship between activated H$_2$ bond length and adsorption distance on MXenes identified with graph neural network and resonating valence bond theory

Jiewei Cheng[a], Tingwei Li[a], Yongyi Wang[b], Ahmed H. Ati[a] and Qiang Sun[a, c*]

[a]School of Materials Science and Engineering, Peking University, Beijing 100871, China
[b]College of Engineering, Peking University, Beijing 100871, China
[c]Center for Applied Physics and Technology, Peking University, Beijing 100871, China

**ABSTRACT:** Motivated by the recent experimental study on hydrogen storage in MXene multilayers [*Nature Nanotechnol*. 2021, 16, 331], for the first time we propose a workflow to computationally screen 23,857 compounds of MXene to explore the general relation between the activated H$_2$ bond length and adsorption distance. By using density functional theory (DFT), we generate a dataset to investigate the adsorption geometries of hydrogen on MXenes, based on which we train physics-informed atomistic line graph neural networks (ALIGNNs) to predict adsorption parameters. To fit the results, we further derived a formula that quantitatively reproduces the dependence of H$_2$ bond length on the adsorption distance from MXenes within the framework of Pauling's resonating valence bond (RVB) theory, revealing the impact of transition metal's ligancy and valence on activating dihydrogen in H$_2$ storage.

## Introduction

One of the most promising solutions to achieving carbon neutrality is to use hydrogen energy, which has renewability and zero CO$_2$ emission[1-3]. However, the most difficult challenge is to find materials that can store hydrogen with large gravimetric and volumetric density and operate under ambient thermodynamic conditions[4-6]. While the bonding of hydrogen existed in nature is either too strong as in metal hydrides, or too weak as in MOFs[7]. To balance the thermodynamics and kinetics, hydrogen binding needs to be between physisorption and chemisorption, namely, in quasi-molecular form, where H$_2$ is activated with elongated H-H bond length.

The activation can be achieved either by electron transferring like in Kubas effect[8] or by charge polarization unveiled by Jena and co-workers[9]. The former takes advantage of the unfilled *d* orbitals of transition-metal atoms where H$_2$ molecules donate electrons to the unfilled *d* orbitals and the transition metals back donate the electrons to the H$_2$ molecules, and the latter relies on the charge polarization induced by the local electrical field. In both cases, H$_2$ retains its molecular bond but with stretched H-H bond length.

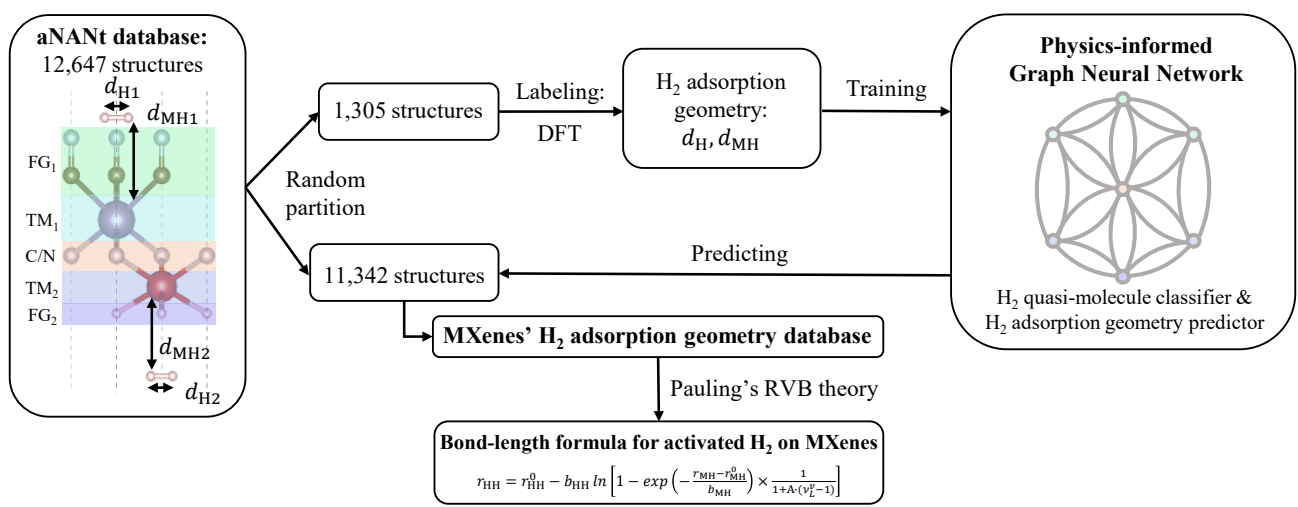

Figure 1. Workflow for the high-throughput screening of MXene for hydrogen storage via graph neural network and multiscale simulation.

The H-H bond length is like a barometer that directly measures the interaction strength of $H_2$ molecule with materials, while the interaction directly depends on the distance of $H_2$ from the substrates. Therefore, it is of vital significance to explore the relation of H-H bond length changing with the adsorption distance, which motivates the studies in this direction. For example, Gründemann et al.[10] developed an equation based on Pauling's bond order concept and experimental evidences to explain the H-H bond length and metal-H distances' correlations in transition metal dihydrides and dihydrogen complexes. Zhou et al.[11] computationally studied external electric field's impact on stretching dihydrogen bond and adsorption distance on boron nitride sheet. Friederich et al.[12] revealed the correlation between H-H bond length and activation process of dihydrogen in Vaska's complexes via *ab-initio* simulations and machine learning. These studies clearly showed that the pursued relationship depends on the substrate materials.

Recently, MXenes have been found to be promising candidates both as hydrogen storage materials[13-15] and as catalysts for enhancing metal hydrides' performance[16-18] due to its superior ability in activating $H_2$ bonds. However, current studies have only covered a very small part of the MXene compounds and the underlying hydrogen activation mechanisms on MXenes are still unclear. Intriguingly, the transition metal (TM) sites are normally coordinated in MXenes, while one functional group is bonded to three TM sites. Therefore, the dihydrogen bond activation process is neither on open metal sites nor on normally saturated TM sites. Two-dimensional (2D) MXene monolayers are building blocks of multilayer MXenes. It is of importance to explore the activation of dihydrogen on 2D MXene layers, especially, the general relationship between H-H bond length and the adsorption distance from MXenes.

In this work, we design a workflow by combining high-throughput simulation and machine learning to screen the MXenes' material space for hydrogen activation, as shown in Figure 1, where the state-of-art ALIGNN model[19] and physics-informed machine learning[20] are applied.

**Results and discussion**

**MXene's hydrogen adsorption geometry datasets**

According to experiment[14] and aNANt database[21], the MXenes' $H_2$ adsorption geometry dataset is constructed.

MXenes' $H_2$ adsorption geometry refers to $H_2$'s bond length and its adsorption distance from MXene. In a unit cell of MXene (see Figure 1), five joint layers are arranged in an array of $FG_1$, $TM_1$, C/N, $TM_2$ and $FG_2$ ($FG_{1/2}$: functional group, $TM_{1/2}$: early transition metal, C/N: carbon or nitrogen). According to aNANt database[21], there are 11 and 13 choices for TM sites and FG layers respectively, which eventually generates 23,857 distinct computationally designed 2D MXenes monolayers. To avoid jeopardizing hydrogen storage performance, MXenes with fifth period transition metals (i.e. Hf, Ta and W) are excluded due to their heavy atomic weight. The remaining 12,647 structures are randomly divided into 1,305 and 11,342 subsets (Figure S2) as the training and prediction subsets. We assign one hydrogen molecule on each TM site and conducted high-throughput calculation to obtain hydrogen adsorption geometry for each MXene monolayer in the first subset. More details of our simulation can be found in Supplementary Information.

We then extract four labels to identify the adsorption geometry of $H_2$ on MXenes: the bond lengths of two hydrogen molecules ($d_{H1}$, $d_{H2}$), the distances between TM sites and geometry centers of hydrogen molecules ($d_{MH1}$, $d_{MH2}$). The $H_2$ bond length and adsorption distance can faithfully reflect the process of $H_2$'s adsorption, as is illustrated in Figure 2. A potential energy surface scan is performed to $H_2$ adsorbed on a $3 \times 3 \times 1$ supercell of $Ti_2CH_2$ (Figure 2(a-b)), where $H_2$'s z coordinate and sorbent atoms are fixed and $H_2$'s bond length is able to relax. Figure 2(c) shows that when $H_2$ approaches the sorbent, the $d_H$ scarcely stretches and adsorption energy $E_{ads}$ increases slightly due to charge polarization. When $d_{MH}$ is smaller than 4 Å, H-H bond and $E_{ads}$ observably increases with the decrease of adsorption distance primarily due to Kubas effect. Therefore, the four adsorption geometry labels are sufficient to give an insight into H$_2$'s asorption on MXenes.

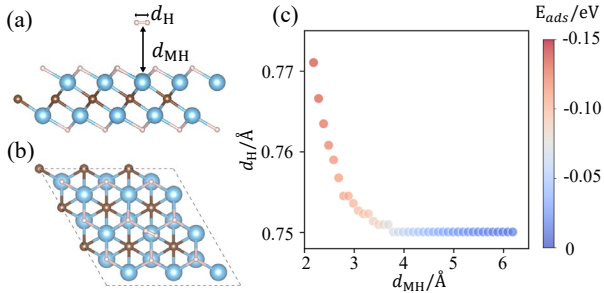

Figure 2. (a) side view and (b) top view of $H_2$ adsorbed on a $3 \times 3 \times 1$ supercell of $Ti_2CH_2$; (c) adsorption energy of $H_2$ changing with $d_{MH}$ and $d_H$ on $Ti_2CH_2$.

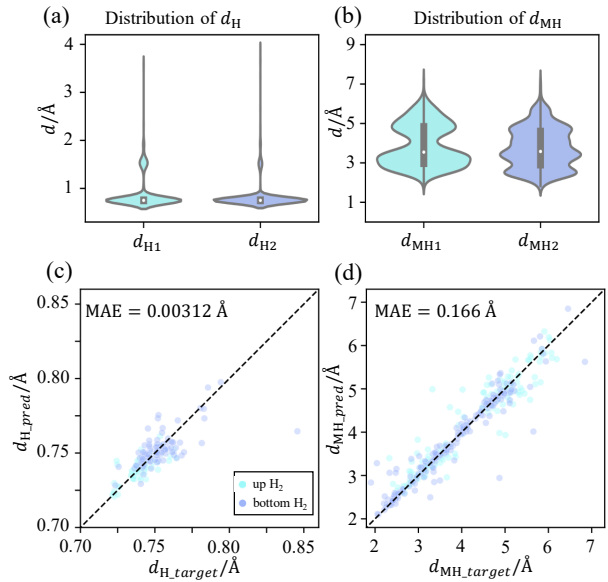

Figure 3. Distribution of distances between (a) two hydrogen atoms and (b) TM and center of $H_2$. PI-ALIGNN predictions versus DFT calculation results of (c) $d_H$ and (d) $d_{MH}$ in the test set. The data related to up/bottom hydrogen molecule are plotted with aqua/light-blue color.

The distributions of these four labels in our dataset are plotted in Figure 3 (a-b) and the relation between ($d_{H1}$, $d_{H2}$) and ($d_{MH1}$, $d_{MH2}$) is plotted in Figure S1. Figure 3 (a) shows that the molecular or quasi-molecular adsorption of H$_2$ happens with $d_H$ close to the equilibrium distance and atomic H binding also happens.

### Two-step ALIGNN model

The first ALIGNN model as H$_2$ quasi-molecule classifier is designed to predict whether the hydrogen maintains a molecule or quasi-molecule form, since the adsorbed H$_2$ can be dissociated in many ways (Figure S3), which will reduce the accuracy of H$_2$ adsorption geometry predictor due to limited amount of data. The model is composed of two identical ALIGNNs in the classification form to predict the state of one hydrogen molecule. If the distance between hydrogen atoms in the adsorbed molecule is smaller than 0.9 Å (1.2 times of the equilibrium distance), the MXene structure will be labeled quasi-molecular configuration. The precision of our two models on test set reaches 0.938 and 0.957, respectively (Table S2).

Next, 1,021 structures in our dataset with two quasi-molecule labels are picked out to train the second model, a H$_2$ adsorption geometry predictor. This model is a multi-output regression ALIGNN trained to predict four hydrogen adsorption geometry labels, namely two H$_2$ bond lengths and two H$_2$ adsorption distances. To achieve a better performance, we inform our model of the fact that when H$_2$ is far away from TM, it tends to maintain its equilibrium bond length by adding a bond-order term in loss function, making it a physics-informed (PI) version. Compared to normal ALIGNN, our PI-ALIGNN's mean MAE of 5 folds is reduced by 0.00013 Å, which is about 4% of total MAE (Table S3-4). The mean MAE of 0.00295 Å in 5-fold cross-validation is in the same order of magnitude as the error range in DFT calculations for geometry (around 0.002 Å), making our model reliable in predicting $d_H$. As shown in Figure 3(c-d), the adsorption geometries of the H$_2$ are well predicted with a MAE of 0.00312 Å for $d_H$ and 0.166 Å for $d_{MH}$ in test set. More details of ALIGNN model's implementation are given in Supporting Information.

Then, this two-step model is applied to the remaining database. The first H$_2$ quasi-molecule classifier ALIGNN picked out 9,883 structures with two quasi-molecule labels from 12,647 pieces of data. The H$_2$ adsorption geometry predictor PI-ALIGNN predicted four hydrogen adsorption geometry labels of these 9,883 MXenes.

### Analysis to MXene's hydrogen storage mechanism

To understand the mechanism of hydrogen adsorption on MXenes both qualitatively and quantitively, we choose the five mono-component FGs, i.e. H, O, F, Cl and Br, which have relatively simple FG-H$_2$ interaction, to analyze the dependence of H$_2$ bond length on adsorption distance (Figure 4). The rest data containing binary and ternary component FGs are presented in Figure S4. Note that we don't discriminate up and bottom H$_2$ in the following discussions.

In the existing research literatures, it has been well accepted that the $d_H$ increases as $d_{MH}$ reduces due to the electron donation and back-donation between the transition metal's $d$ orbital and H$_2$'s $\sigma^*$ orbital. Such picture gives rise to a natural assumption that the total valence bond order of hydrogen atom is unity (since each H has one electron) and can be composed of individual valence bond orders, namely, we have the following equation[10],

$$p_{HH} + p_{MH} = 1 \quad (1)$$

where $p_{HH}$ and $p_{MH}$, the bond order between H-H and TM-H, is given by[22],

$$p_{XH} = exp\left(-\frac{r_{XH} - r_{XH}^0}{b_{XH}}\right) \quad (2)$$

where X is H/M, $r_{XH}$ is the distance between X and H atom, $r_{XH}^0$ represents the equilibrium X-H distance of free donor XH (when $p_{XH} = 1$) and $b_{XH}$ is a decay parameter. A quantitative relation reads[10],

$$r_{HH} = r_{HH}^0 - b_{HH} ln\left[1 - exp\left(-\frac{r_{MH} - r_{MH}^0}{b_{MH}}\right)\right] \quad (3)$$

**Table 1. Different values of $v_L^v$ for TMs in MXenes**

| element (ligancy 6) | Sc | Ti | V | Cr | Y | Zr | Nb | Mo |
|---|---|---|---|---|---|---|---|---|
| valence number $v$ | 3 | 4 | 5 | 6 | 3 | 4 | 5 | 6 |
| $v_L^v$ | 6.25 | 5.48 | 3.62 | 1.86 | 6.25 | 5.48 | 3.62 | 1.86 |

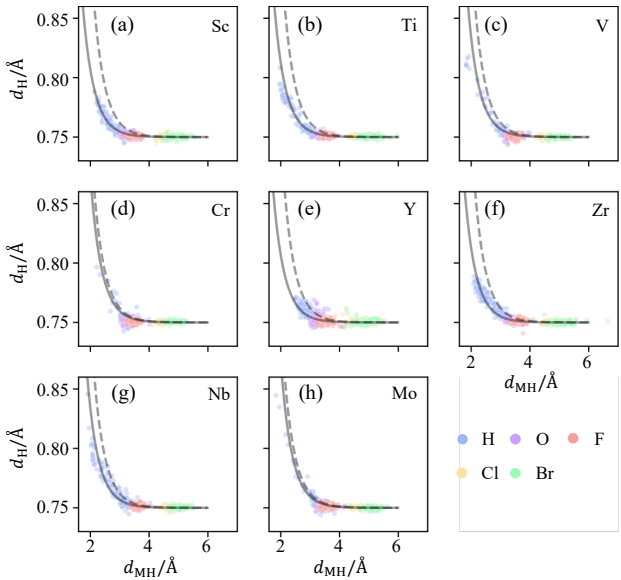

Figure 4. Predicted along with calculated hydrogen adsorption geometries of the five mono-component FGs separated by different TMs with (a) Sc, (b) Ti, (c) V, (d) Cr, (e) Y, (f) Zr, (g) Nb and (h) Mo. Dashed lines are from Eq. (3) and solid lines from Eq. (6), our formula for H$_2$ adsorption geometry.

where we set $r_{MH}^0 = 1.6$ Å and $b_H = b_{MH} = 0.404$ Å according to the literature[10] and equilibrium $H_2$ bond length $r_{HH}^0 = 0.75$ Å in line with our computational result. Given that

$$r_{HH} = d_H, \quad r_{MH} = \sqrt{d_{MH}^2 + \left(\frac{d_H}{2}\right)^2}, \quad (4)$$

The dashed lines in Figure 4 are from Eq. (3), suggesting that the Eq. (3) is not suitable for the functionalized MXenes. It was originally derived for open TM sites in complexes and TM hydrides, which overestimates $d_H$ in our systems with normally coordinated metal sites (NCMSs). TM sites in MXenes have a ligancy 6 which lowers their ability to activate dihydrogen bonds. A revised equation is needed for MXenes and other materials by taking ligancy and valence into consideration.

For this purpose, we adapt a new formula for the bond order $p_{MH}^*$ between TM and H atom, deduced from Pauling's RVB theory through a statistic treatment of valence electrons[23-24],

$$p_{MH}^* = exp\left(-\frac{r_{MH}-r_{MH}^0}{b_{MH}}\right) \times \frac{1}{1+A\cdot(v_L^v-1)} \quad (5)$$

where A is a coupling coefficient of resonating between different structures and $v_L^v$ is the number of unsynchronized resonance structures per atom, determined with valence $v$ and ligancy $L$, given by Eq. (S4) (see Supporting Information for detailed deductions). Eq. (5) gives a basic picture of the resonating process of valence bonds. Strength of resonance is represented with coupling constant A and possible resonating structures $v_L^v$. The valence electrons are delocalized due to resonance so that the bond strength, or bond order, decreases under the same bond distance $r_{MH}$. The bond order of dihydrogen bond is still described with Eq. (2) since $H_2$ has no ligand.

The values of $v_L^v$ for TMs in this work is shown in Table 1, and the revised formula for $H_2$ adsorption geometry on NCMS reads

$$r_{HH} = r_{HH}^0 - b_{HH} ln\left[1 - exp\left(-\frac{r_{MH}-r_{MH}^0}{b_{MH}}\right) \times \frac{1}{1+A\cdot(v_L^v-1)}\right]. \quad (6)$$

We choose A = 0.3 to achieve a good agreement with our data (solid lines in Figure 4). Through $v_L^v$, the activation of $H_2$ bond on MXenes is properly described by Eq. (6) as the solid lines plotted in Fig.3. What's more, TM with more valence electrons tends to have longer H-H bond length at the same adsorption distance, implying a better activation of H-H bond. From the perspective of Pauling's RVB theory, the increase of $v_L^v$ for the unsynchronized resonance structures per atom results from the lack of valence electron compared with atom's ligancy. The resonance between different electron configurations delocalizes the electrons, thus TM's ability to activate $H_2$ bond is weakened. Therefore, TM with more valence electrons tends to perform as better dihydrogen bond activator in MXenes, whose TM sites all have same ligancy of 6.

Eq. (6) gives us a quantitative correlation between dihydrogen bond length and $H_2$'s adsorption distance from NCMSs on MXenes with mono-component FGs, which can be readily extended to other crystal material surfaces with similar local structures.

**Conclusion**

In summary, to better understand the recent experiment on hydrogen storage in multilayer MXenes, we focus on the activation of $H_2$ on the functionalized MXenes, which can be measured by the H-H bond length changing with the adsorption distance. Based on high-throughput simulation, machine learning and Pauling's RVB theory, we derived a general formula for describing the relationship between activated $H_2$ bond length and adsorption distance, which could be extended to other 2D materials and materials surfaces to get insight on $H_2$'s activation process.

ASSOCIATED CONTENT

The implementation and detailed analysis of high-throughput simulation, ALIGNN model and related deductions in Pauling's RVB theory are available in the Supporting Information. This material is available free of charge via the Internet at http://pubs.acs.org.

AUTHOR INFORMATION

Corresponding Author

* Qiang Sun - School of Materials Science and Engineering, Peking University, Beijing 100871, China; Center for Applied Physics and Technology, Peking University, Beijing 100871, China; Email: sunqiang@pku.edu.cn

ACKNOWLEDGMENT

This work was partially supported by grants from the National Key Research and Development Program of China (2021YFB4000601), and from the China Scholarship Council (CSC). The calculations were supported by the High-performance Computing Platform of Peking University.

REFERENCES

1. Jena, P., Materials for Hydrogen Storage: Past, Present, and Future. The Journal of Physical Chemistry Letters 2011, 2 (3), 206-211.
2. Chu, S.; Majumdar, A., Opportunities and challenges for a sustainable energy future. Nature 2012, 488 (7411), 294-303.
3. Yang, X.; Nielsen, C. P.; Song, S.; McElroy, M. B., Breaking the hard-to-abate bottleneck in China's path to carbon neutrality with clean hydrogen. Nature Energy 2022, 7 (10), 955-965.
4. Allendorf, M. D.; Stavila, V.; Snider, J. L.; Witman, M.; Bowden, M. E.; Brooks, K.; Tran, B. L.; Autrey, T., Challenges to developing materials for the transport and storage of hydrogen. Nature Chemistry 2022, 14 (11), 1214-1223.
5. Kubas, G. J., Breaking the H$_2$ Marriage and Reuniting the Couple. Science 2006, 314 (5802), 1096-1097.
6. Meduri, S.; Nandanavanam, J., Materials for hydrogen storage at room temperature–An overview. Materials Today: Proceedings 2022.
7. Chen, Z.; Kirlikovali, K. O.; Idrees, K. B.; Wasson, M. C.; Farha, O. K., Porous materials for hydrogen storage. Chem 2022, 8 (3), 693-716.
8. Kubas, G. J., Metal–dihydrogen and σ-bond coordination: the consummate extension of the Dewar–Chatt–Duncanson model for metal–olefin π bonding. Journal of Organometallic Chemistry 2001, 635 (1-2), 37-68.
9. Niu, J.; Rao, B. K.; Jena, P., Binding of hydrogen molecules by a transition-metal ion. Physical Review Letters 1992, 68 (15), 2277-2280.

10. Gründemann, S.; Limbach, H.-H.; Buntkowsky, G.; Sabo-Etienne, S.; Chaudret, B., Distance and Scalar HH-Coupling Correlations in Transition Metal Dihydrides and Dihydrogen Complexes. The Journal of Physical Chemistry A 1999, 103 (24), 4752-4754.

11. Zhou, J.; Wang, Q.; Sun, Q.; Jena, P.; Chen, X. S., Electric field enhanced hydrogen storage on polarizable materials substrates. Proceedings of the National Academy of Sciences 2010, 107 (7), 2801-2806.

12. Friederich, P.; dos Passos Gomes, G.; De Bin, R.; Aspuru-Guzik, A.; Balcells, D., Machine learning dihydrogen activation in the chemical space surrounding Vaska's complex. Chemical Science 2020, 11 (18), 4584-4601.

13. Kumar, P.; Singh, S.; Hashmi, S. A. R.; Kim, K.-H., MXenes: Emerging 2D materials for hydrogen storage. Nano Energy 2021, 85, 105989.

14. Liu, S.; Liu, J.; Liu, X.; Shang, J.; Xu, L.; Yu, R.; Shui, J., Hydrogen storage in incompletely etched multilayer Ti2CTx at room temperature. Nature Nanotechnology 2021, 16 (3), 331-336.

15. Ghotia, S.; Kumar, A.; Sudarsan, V.; Dwivedi, N.; Singh, S.; Kumar, P., Multilayered Ti3C2Tx MXenes: A prominent materials for hydrogen storage. International Journal of Hydrogen Energy 2023.

16. Huang, T.; Huang, X.; Hu, C.; Wang, J.; Liu, H.; Xu, H.; Sun, F.; Ma, Z.; Zou, J.; Ding, W., MOF-derived Ni nanoparticles dispersed on monolayer MXene as catalyst for improved hydrogen storage kinetics of MgH2. Chemical Engineering Journal 2021, 421, 127851.

17. Zhu, W.; Ren, L.; Lu, C.; Xu, H.; Sun, F.; Ma, Z.; Zou, J., Nanoconfined and in Situ Catalyzed MgH2 Self-Assembled on 3D Ti3C2 MXene Folded Nanosheets with Enhanced Hydrogen Sorption Performances. ACS Nano 2021, 15 (11), 18494-18504.

18. Liu, H.; Duan, X.; Wu, Z.; Luo, H.; Wang, X.; Huang, C.; Lan, Z.; Zhou, W.; Guo, J.; Ismail, M., Exfoliation of compact layered Ti2VAlC2 MAX to open layered Ti2VC2 MXene towards enhancing the hydrogen storage properties of MgH2. Chemical Engineering Journal 2023, 468, 143688.

19. Choudhary, K.; DeCost, B., Atomistic Line Graph Neural Network for improved materials property predictions. npj Computational Materials 2021, 7 (1), 185.

20. Karniadakis, G. E.; Kevrekidis, I. G.; Lu, L.; Perdikaris, P.; Wang, S.; Yang, L., Physics-informed machine learning. Nature Reviews Physics 2021, 3 (6), 422-440.

21. Rajan, A. C.; Mishra, A.; Satsangi, S.; Vaish, R.; Mizuseki, H.; Lee, K.-R.; Singh, A. K., Machine-Learning-Assisted Accurate Band Gap Predictions of Functionalized MXene. Chemistry of Materials 2018, 30 (12), 4031-4038.

22. Pauling, L., Atomic Radii and Interatomic Distances in Metals. Journal of the American Chemical Society 1947, 69 (3), 542-553.

23. Pauling, L., The metallic orbital and the nature of metals. Journal of Solid State Chemistry 1984, 54 (3), 297-307.

24. Pauling, L.; Kamb, B., A revised set of values of single-bond radii derived from the observed interatomic distances in metals by correction for bond number and resonance energy. Proceedings of the National Academy of Sciences 1986, 83 (11), 3569-3571.

TOC Graphic

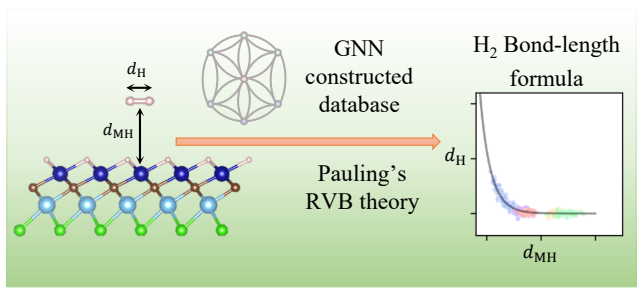